\documentclass[journal]{IEEEtran}
\usepackage{graphicx}
\usepackage{amsmath}
\usepackage{subfigure}
\usepackage{multirow}
\usepackage{array}

\ifCLASSINFOpdf

\else

\fi

\begin{document}
\title{DSDV, DYMO, OLSR:\\ Link Duration and Path Stability}

\author{\IEEEauthorblockN{S. Kumar, N. Javaid, Z. Yousuf, H. Kumar, Z. A. Khan$^{\S}$, A. Bibi\\\vspace{0.4cm}}
                Department of Electrical Engineering, COMSATS\\ Institute of
                Information Technology, Islamabad, Pakistan. \\
                $^{\S}$Faculty of Engineering, Dalhousie University, Halifax, Canada.
             }

%
%% make the title area

\maketitle

\begin{abstract}
In this paper, we evaluate and compare the impact of link duration and path stability of routing protocols; Destination Sequence Distance vector (DSDV), Dynamic MANET On-Demand (DYMO) and Optimized Link State Routing (OLSR) at different number of connections and node density. In order to improve the efficiency of selected protocols; we enhance DYMO and OLSR. Simulation and comparison of both default and enhanced routing protocols is carried out under the performance parameters; Packet Delivery Ratio (PDR), Average End-to End Delay (AE2ED) and Normalized Routing Overhead (NRO). From the results, we observe that DYMO performs better than DSDV, MOD-OLSR and OLSR in terms of PDR, AE2ED, link duration and path stability at the cost of high value of NRO.
\end{abstract}


\IEEEpeerreviewmaketitle

%section1
\section{Introduction}
Mobile Ad-hoc Networks (MANETs) are dynamic and self-organized networks and do not require any prefixed infrastructure. In these networks, nodes are mobile and act like routers to communicate with each other. Vehicular Ad-hoc Networks (VANETs) are the special case of MANETs, in which mobile nodes are vehicles with radio communication range of 250 to 300 meters [1]. In VANETs, nodes have high mobility that causes fast change of the topology, therefore, its link stability is less than MANETs. Speeds of vehicles moving in same direction are similar most of the time, therefore, they remain in radio contact for longer time than vehicles moving in opposite direction. So, path stability in VANETs depends on vehicle density and number of connection between these vehicles. When vehicle density and number of connections are less, then link breakage will be more and therefore, link stability decreases and Normalized Routing Overhead (NRO) increases. As far as safety is concerned, VANETs are more appropriate and reliable for this purpose because they exhibit road accidents and traffic jams.

This paper discusses the performance of proactive and reactive routing protocols in accordance of performance parameters for a urban scenario in VANETs. Nakagami model is used for simulation work in NS-2 because in [2], authors conclude that Nakagami model experimentally performs well among the available propagation models. We simulate and analyze both default and modified routing protocols; DSDV [3], DYMO [4] and OLSR [5] under the performance parameters; Packet Delivery Rate (PDR), Average End-to-End Delay (AE2ED) and NRO. To improve efficiency we enhance DYMO and OLSR. In DYMO, $network~diameter$ from $10$ to $30$ hops and $Route~Request~wait~time$ from $1000$ to $600$ seconds. While in OLSR, $Topology~Control~(TC)~Message~interval$ from $5$ to $3$ seconds and $Hello~Message~interval$ from $2$ to $1$ seconds. Through these modifications, comparison of routing protocols; DYMO and OLSR is carried out in accordance of performance parameters.

Rest of the paper is organized as: Related Work and Motivation discussed in section II. Section III finds the link duration and path stability of vehicles in different routing protocols. In section IV, the simulation results ar discussed. Performance trade-off is explained in section V. Conclusion of results is analyzed in section VI.

\section{Related Work and Motivation}

\subsection{Related Work}
In [6], authors evaluate three routing protocols; selected from categories; geographic routing (i.e., GPSR), geographic opportunistic routing (i.e., GOSR), and trajectory based routing (i.e., SIFT) for VANETs in urban environments. In order to model realistic vehicular pattern, Vehicular Mobility Model (VMM) is used. They analyzed routing protocols varying vehicle density and speed against the performance parameters; PDR,  Packet Loss Ratio (PLR), Throuput, AE2ED, Average number of hops and control overhead.

[7] presents the evaluation of IEEE 802.11p with IEEE 802.11a. Simulations are performed in NS-2 using two MAC protocols; 802.11a and 802.11p. Three performance parameters are measured; AE2ED, throughput, packet drops during various modes. From the observed results, it is concluded that 802.11p performs better than 802.11a while considering different performance parameters.

Authors in [8], predict the link duration and stability of nodes in MANETs. They find link duration (for how much time link is available between nodes) of nodes and also calculated the mean duration. On the basis of link duration, they found link stability of nodes keeping one node at fixed position while other is moving with relative velocity.

In [9], authors revealed the prediction of link stability through the changes in link connectivity. Further comparing the link connectivity, based prediction schemes with other papers. They proposed a scheme, derived analytically using a probabilistic model in MANETs.

\subsection{Motivation}
Motivation is taken from the papers as mentioned in related work and also from simulation results discussed in section IV. In this paper we have done simulation in urban scenario that was due to motivated by [6-9]. In [6], routing protocols are evaluated in VANETs for urban scenario. Paper [7] evaluate and compare MAC protocols; IEEE 802.11p and IEEE 802.11a under performance parameters. Authors in [8], find link duration ($D_{oa}$) of link ($L_{oa}$) between two nodes $A$ and $O$ from current time $t$ to time at which $L_{ao}$ is broken, keeping one node at fixed position, while other is moving. [9] presents the probabilistic model for link stability through the changes in link connectivity. Inspired by [6] and [7], we take urban scenario for simulation work using IEEE 802.11p as good one. We evaluate and compare the performance of three routing protocols; DSDV, DYMO and OLSR with four different cases of angles between vehicles. Further, link duration and path stability is determined between vehicles being motivated by [8] and [9].

\section{Link Duration and Path Stability Modling}
In [8], authors find link duration $(D_{oa})$ of link $(L_{oa})$ between two nodes $A$ and $O$ from current time $t$ to time at which $L_{ao}$ is broken. They take two nodes as mobile, keeping one node at fixed position while other is moving. Further they calculate mean link duration $({\overline{D}_{oa}})$ based on distance $(d)$.

We consider urban scenario in VANETs, in which nodes (vehicles) are moving with different velocities. The links between vehicles are not available for longer time and the relative velocity (Difference of velocities between two vehicles and expressed as $v_r$) of vehicles is changed at different time instants. Path stability depends on available links between the vehicles, so their stability is also decreased. Therefore, we find link duration and path stability between vehicles for four cases. These cases are discussed below:

\subsection{Case-I}
In this case, It is assumed that at time $t_0$ the distance between two vehicles $A_{t_0}$ and $B_{t_0}$ is $d_{t_0}$. At time $t_1$, $A_{t_0}$ and $B_{t_0}$ move with distances $d_1$ and $d_2$ making angles of $\alpha_{O}$ and $\beta_{O}$, respectively, as shown in Fig. 1. Further we calculate the distances $R_1$ and $R_2$ between $A_{t_0}$ and $B_{t_1}$, and $B_{t_0}$ and $A_{t_1}$, using cosine law and angles $\Psi_A$ and $\Psi_B$ using sine law, respectively, given in Eq. (1,3,2 and 4).
\begin{figure}[h]
  \centering
  {\includegraphics[height=8 cm,width=8 cm]{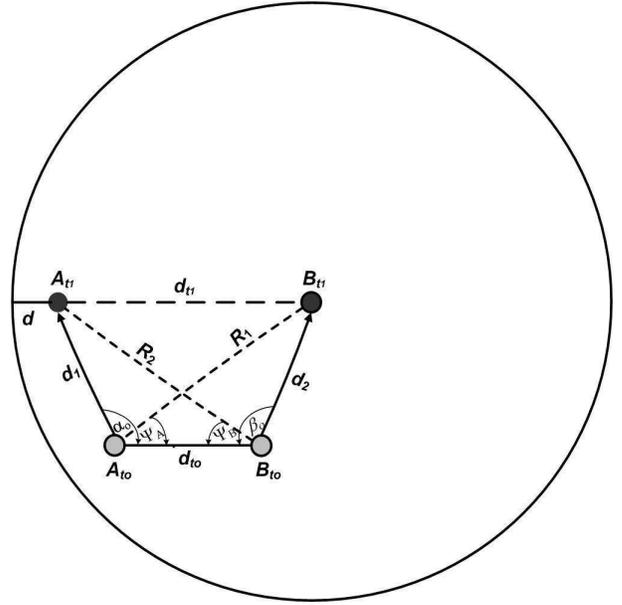}}
  \caption{Vehicles are moving with obtuse angles}
\end{figure}
\begin{eqnarray}
 R_{1} &=& \sqrt{{d_{t_0}}^2+{d_{2}}^2-2{d_{t_0}}{d_{2}}\cos{(\beta_{obtuse})}}
\end{eqnarray}
\begin{eqnarray}
  \Psi_A &=& \arcsin{(\frac{d_2\sin{(\beta_{obtuse})}}{R_1})}
\end{eqnarray}
\begin{eqnarray}
 R_{2} &=& \sqrt{{d_{t_0}}^2+{d_{1}}^2-2{d_{t_0}}{d_{1}}\cos{(\alpha_{obtuse})}}
\end{eqnarray}
\begin{eqnarray}
  \Psi_B &=& \arcsin{(\frac{d_1\sin{(\alpha_{obtuse})}}{R_2})}
\end{eqnarray}
Then we determine the distance $d_{t_1}$ between $A_{t_1}$ and $B_{t_1}$ at time $(t_1)$ as
\begin{eqnarray}
 d_{t_1} &=& \sqrt{{d_{1}}^2+{R_{1}}^2-2{d_{1}}{R_{1}}\cos{(\alpha_{obtuse}-\Psi_A)}}
\end{eqnarray}
or
\begin{eqnarray}
 d_{t_1} &=& \sqrt{{d_{2}}^2+{R_{2}}^2-2{d_{2}}{R_{2}}\cos{(\beta_{obtuse}-\Psi_B)}}
\end{eqnarray}
or
\begin{eqnarray}
\begin{split}
  d_{t_1} =  d_{t_0}+d_{1}\cos{(180^{\circ}-\alpha_{obtuse})}\\+d_{2}\cos{(180^{\circ}-\beta_{obtuse})}
\end{split}
\end{eqnarray}
Let $r$ be the radio communication range of any node, therefore, the distance $d_{t_1}~\leq~r$. So, the distance $d$ at time $t_1$ will be expressed as
\begin{eqnarray}
  d &=&  r-d_{t_1}
\end{eqnarray}
It is clear that link availability between vehicles depends on two parameters; distance $(d_{t_1})$ and relative velocity $v_r$, therefore in order to find link duration $LD$, we derive an expression as:
\begin{eqnarray}
 LD_{AB} &=& \frac{d}{v_r}
\end{eqnarray}
If the link duration increases, path stability becomes high.

\subsection{Case-II}
In the first case, vehicles are moving with making obtuse angles i.e., greater than $90^\circ$. Here the case is different due to the movement of vehicles with acute angles i.e., less than $90^\circ$. For this case, equations that we drove for case-1 will be same with little change of angles $\alpha_{O}$ as $\alpha_{A}$ and $\beta_{O}$ as $\beta_{A}$ and also the angles $\Psi_A$ and $\Psi_B$ will change due acute angle, as shown in Fig. 2(a) below and also in Eq. (2 and 4). Then, the distances $R_1$ and $R_2$ and the angles $\Psi_A$ and $\Psi_B$ are calculated same like case-1, to find the distance $d_{t_1}$ between $A_{t_1}$ and $B_{t_1}$ at time $t_1$. After that, link duration and path stability between vehicles and distance $d_{t_1}$ in Eq. (10) are determined.
\begin{figure}[h]
  \centering
 \subfigure[Vehicles are moving with acute angles]{\includegraphics[height=3.25 cm,width=4 cm]{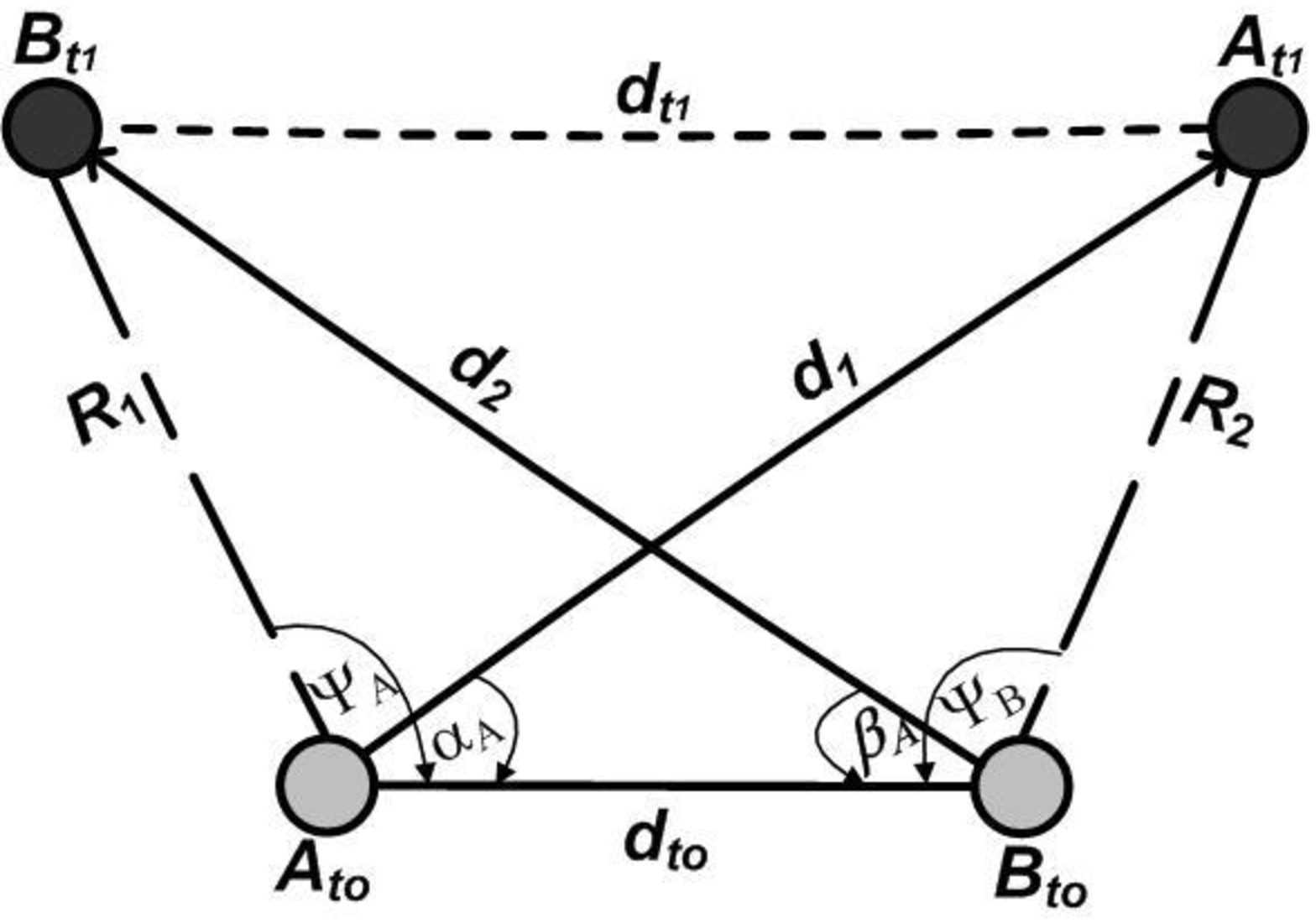}}
 \subfigure[Vehicles are moving with both acute and obtuse angles]{\includegraphics[height=3.25  cm,width=4 cm]{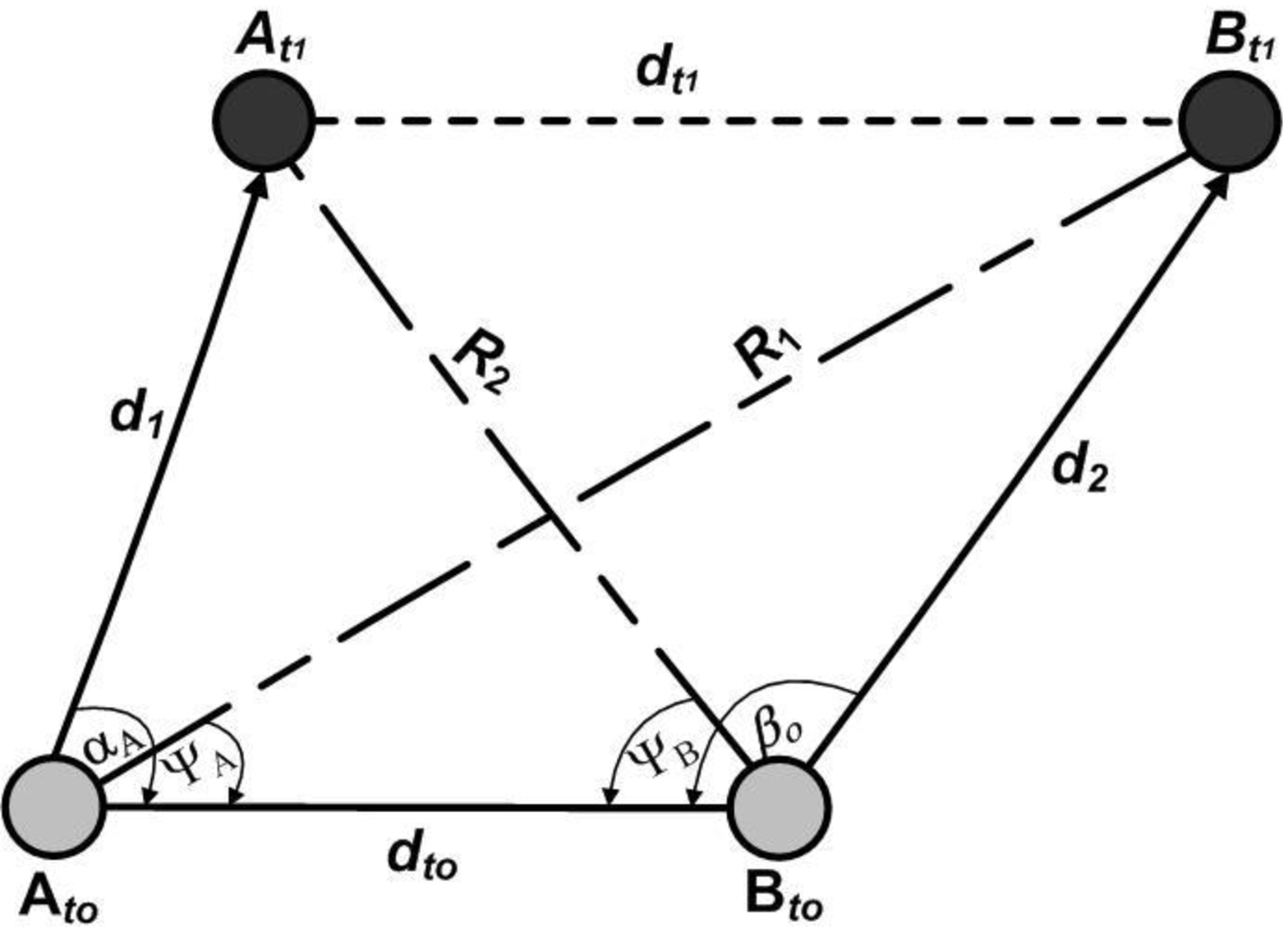}}
 \caption{Moment of vehicles at different angles}
\end{figure}
%$\begin{eqnarray}
    %R_{1} &=& \sqrt{{d_{t_0}}^2+{d_{2}}^2-2{d_{t_0}}{d_{2}}\cos{(\beta_{obtuse})}}
%\end{eqnarray}
%\begin{eqnarray}
   % \Psi_A &=& \arcsin{(\frac{d_2\sin{(\beta_{obtuse})}}{R_1})}
%\end{eqnarray}
%\begin{eqnarray}
    %R_{2} &=& \sqrt{{d_{t_0}}^2+{d_{1}}^2-2{d_{t_0}}{d_{1}}\cos{(\alpha_{obtuse})}}
%\end{eqnarray}
%\begin{eqnarray}
   % \Psi_B &=& \arcsin{(\frac{d_1\sin{(\alpha_{obtuse})}}{R_2})}
%\end{eqnarray}
%\begin{eqnarray}
    %d_{t_1} &=& \sqrt{{d_{1}}^2+{R_{1}}^2-2{d_{1}}{R_{1}}\cos{(\alpha_{obtuse}-\Psi_A)}}
%\end{eqnarray}
%or
%\begin{eqnarray}
    %d_{t_1} &=& \sqrt{{d_{2}}^2+{R_{2}}^2-2{d_{2}}{R_{2}}\cos{(\beta_{obtuse}-\Psi_B)}}
%\end{eqnarray}
\begin{eqnarray}
  d_{t_1} =  d_{t_0}+R_{1}\cos{(180^{\circ}-\Psi_A)}+R_{2}\cos{(180^{\circ}-\Psi_B)}
\end{eqnarray}

\subsection{Case-III}
In this scenario, assumption is taken as one vehicle is moving with distance $d_1$ by making an acute angle $\alpha_{A}$, while other is moving with distance $d_2$ making an obtuse angle $\beta_{O}$. Where the angles $\Psi_A$ and $\Psi_B$ depend on angles $\alpha_{A}$ and $\beta_{O}$, respectively, as shown in Fig. 2(b) and also in Eq. (2 and 4). Now, to calculate the distance $d_{t_1}$ between $A_{t_1}$ and $B_{t_1}$ at time $t_1$, we use same equation of case-1 and also by using Eq. (11). To find link duration and path stability, that will tell us for how much time link will be available between vehicles, Eq. (9) will be used.

%\begin{eqnarray}
    %R_{1} &=& \sqrt{{d_{t_0}}^2+{d_{2}}^2-2{d_{t_0}}{d_{2}}\cos{(\beta_{obtuse})}}
%\end{eqnarray}
%\begin{eqnarray}
    % \Psi_A &=& \arcsin{(\frac{d_2\sin{(\beta_{obtuse})}}{R_1})}
%\end{eqnarray}
%\begin{eqnarray}
    %R_{2} &=& \sqrt{{d_{t_0}}^2+{d_{1}}^2-2{d_{t_0}}{d_{1}}\cos{(\alpha_{obtuse})}}
%\end{eqnarray}
%\begin{eqnarray}
    % \Psi_B &=& \arcsin{(\frac{d_1\sin{(\alpha_{obtuse})}}{R_2})}
%\end{eqnarray}
%\begin{eqnarray}
 %d_{t_1} &=& \sqrt{{d_{1}}^2+{R_{1}}^2-2{d_{1}}{R_{1}}\cos{(\alpha_{obtuse}-\Psi_A)}}
%\end{eqnarray}
%or
%\begin{eqnarray}
 %d_{t_1} &=& \sqrt{{d_{2}}^2+{R_{2}}^2-2{d_{2}}{R_{2}}\cos{(\beta_{obtuse}-\Psi_B)}}
%\end{eqnarray}
%or
\begin{eqnarray}
    d_{t_1} =  d_{t_0}-d_{1}\cos{(\alpha_{acute})}+d_{2}\cos{(180^{\circ}-\beta_{obtuse})}
\end{eqnarray}

\subsection{Case-IV}
This case is same like case-3 but we change the angles $\alpha_{A}$ as $\alpha_{O}$ and $\beta_{O}$ as $\beta_{A}$, so, the angles $\Psi_A$ and $\Psi_B$ will also be changed, as shown in Fig. 3. Then $R_1$ and $R_2$, and $\Psi_A$ and $\Psi_B$, are dtermind using Eq. (1-4), to calculate the distance $d_1$ between vehicles at time $t_1$ and also using Eq. (12) for $d_1$. Further, we find its link duration and path stability using Eq. (9).

\begin{figure}[h]
  \centering
  {\includegraphics[height=3.25 cm,width=4 cm]{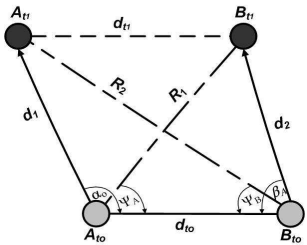}}
  \caption{Vehicles are moving with obtuse and obtuse angles}
\end{figure}
%\begin{eqnarray}
 %R_{1} &=& \sqrt{{d_{t_0}}^2+{d_{2}}^2-2{d_{t_0}}{d_{2}}\cos{(\beta_{obtuse})}}
%\end{eqnarray}
%\begin{eqnarray}
  %\Psi_A &=& \arcsin{(\frac{d_2\sin{(\beta_{obtuse})}}{R_1})}
%\end{eqnarray}
%\begin{eqnarray}
 %R_{2} &=& \sqrt{{d_{t_0}}^2+{d_{1}}^2-2{d_{t_0}}{d_{1}}\cos{(\alpha_{obtuse})}}
%\end{eqnarray}
%\begin{eqnarray}
  %\Psi_B &=& \arcsin{(\frac{d_1\sin{(\alpha_{obtuse})}}{R_2})}
%\end{eqnarray}
%\begin{eqnarray}
% d_{t_1} &=& \sqrt{{d_{1}}^2+{R_{1}}^2-2{d_{1}}{R_{1}}\cos{(\alpha_{obtuse}-\Psi_A)}}
%\end{eqnarray}
%or
%\begin{eqnarray}
% d_{t_1} &=& \sqrt{{d_{2}}^2+{R_{2}}^2-2{d_{2}}{R_{2}}\cos{(\beta_{obtuse}-\Psi_B)}}
%\end{eqnarray}
%or
\begin{eqnarray}
  d_{t_1} =  d_{t_0}+d_{1}\cos{(180^{\circ}-\alpha_{obtuse})}-d_{2}\cos{(\beta_{acute})}
\end{eqnarray}

\section{Experiments and Discussions }
In this paper, Nakagami propagation model in NS- 2.34 is used. The implementation of original version of DSDV is used in NS-2. For implementation of DYMO and OLSR, DYMOUM [10] and OLSR [11] patchs  are used. The map imported in MOVE and scaled down to $4~km$ x $4~km$ in size for reasonable simulation environment. Using MOVE and SUMO, mobility patterns were generated randomly. Table. 1 shows the complete simulation parameters used in this paper.
\begin{table}[h]
\caption{SIMULATION PARAMETERS}
\begin{center}
 \begin{tabular}{| c | c |}
  \hline
  \textbf{Parameters} & \textbf{Values} \\ \hline
    NS-2 Version &  2.34 \\ \hline
    DSDV Implementation &	NS-2 default \\ \hline
    DYMO Implementation	&   DYMOUM-patch [10] \\ \hline
    OLSR Implementation  &	OLSR-patch [11] \\ \hline
    MOVE version  &	  2.81 \\ \hline
    SUMO version  &	  0.12.3 \\ \hline
    Number of nodes & 30, 50, 70, 90, 120 \\ \hline
    Number of CBR sessions & 6, 12, 18, 24 \\ \hline
    Tx Range &	300m \\ \hline
    Simulation Area &	4KM x 4KM \\ \hline
    Speed &	Uniform, 40kph \\ \hline
    Data Type &	CBR \\ \hline
    Data Packet Size &	1000 bytes \\ \hline
    MAC Protocol &	IEEE 802.11 Overhauled \\ \hline
    PHY Standard &	IEEE 802.11p \\ \hline
    Radio Propagation Model &	Nakagami \\ \hline

 \end{tabular}
\end{center}
\end{table}
The following performance parameters are used to evaluate the performance of routing protocols; AODV, DSDV, DSR, DYMO, FSR and OLSR.

\subsection{PDR}
The ratio of data packets at the destination and total data packets generated. Fig. 4 shows PDR against number of connections. From Fig. 4(a), it is clear that MOD-DYMO attains more PDR than other routing protocols; DSDV, DYMO, MOD-OLSR and OLSR, due to reactive in nature because reactive protocols do not need route calculation before data transmission. So, as number of connections increases, it attains higher PDR than other protocols. While DYMO is showing second highest value in PDR, however, as the number of connections increase, its PDR goes down. In low number of connections, DSDV attains high PDR than MOD-OLSR and OLSR, due to generation of more data packets. In high number of connections, there is occurrence of more full dumps and more drop of data packets, will cause more NRO thus PDR decreases. While, MOD-OLSR and OLSR show increasing graph, The main reason of increasing PDR is that the computation of Multipoint Relay (MPRs) mechanism generates more routing packets therefore, its PDR goes up as increase in number of connections.
\begin{figure}[h]
  \centering
 \subfigure[PDR vs Number of connections]{\includegraphics[height=3 cm,width=4.3 cm]{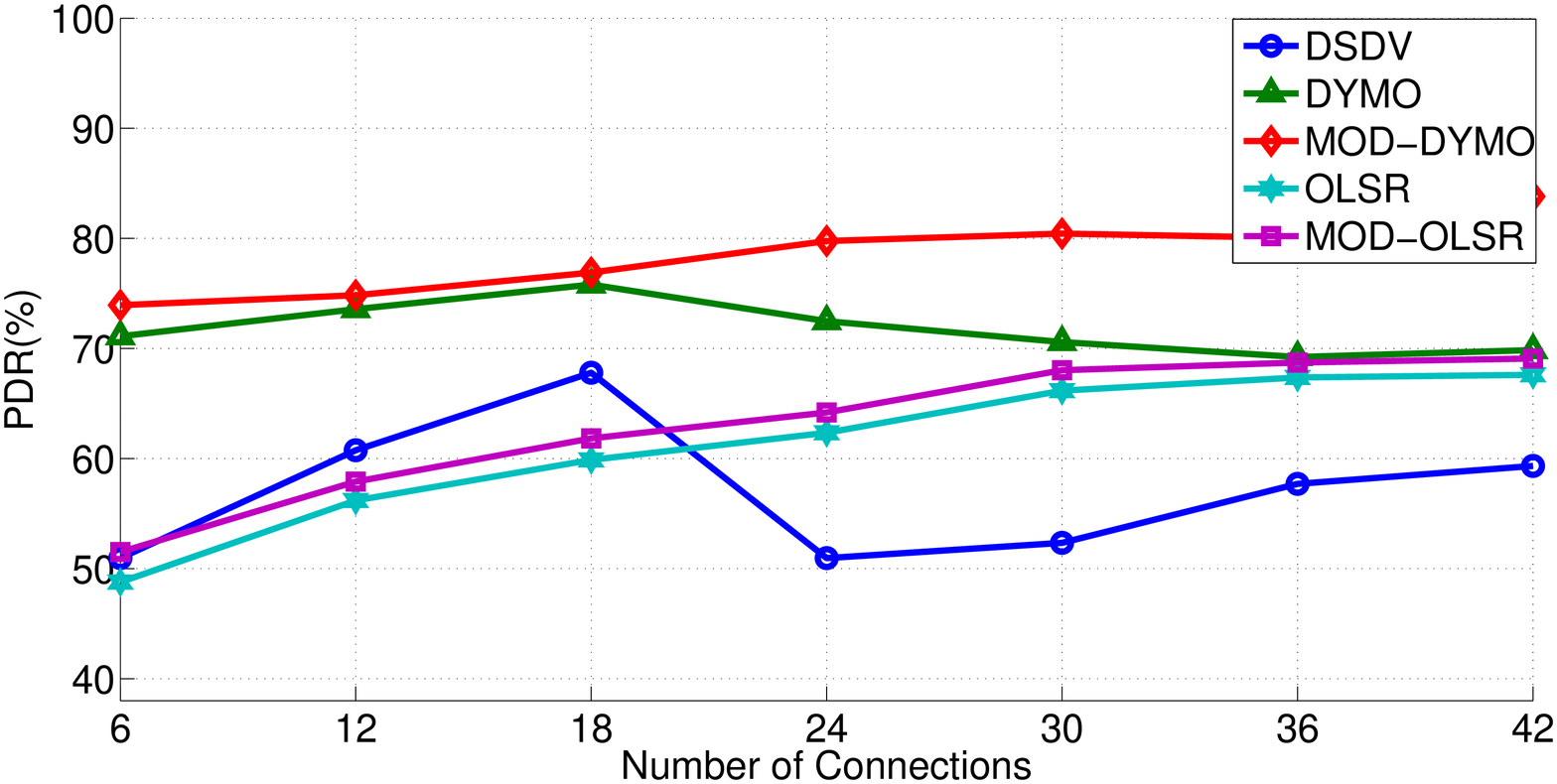}}
 \subfigure[PDR vs Node Density]{\includegraphics[height=3  cm,width=4.3 cm]{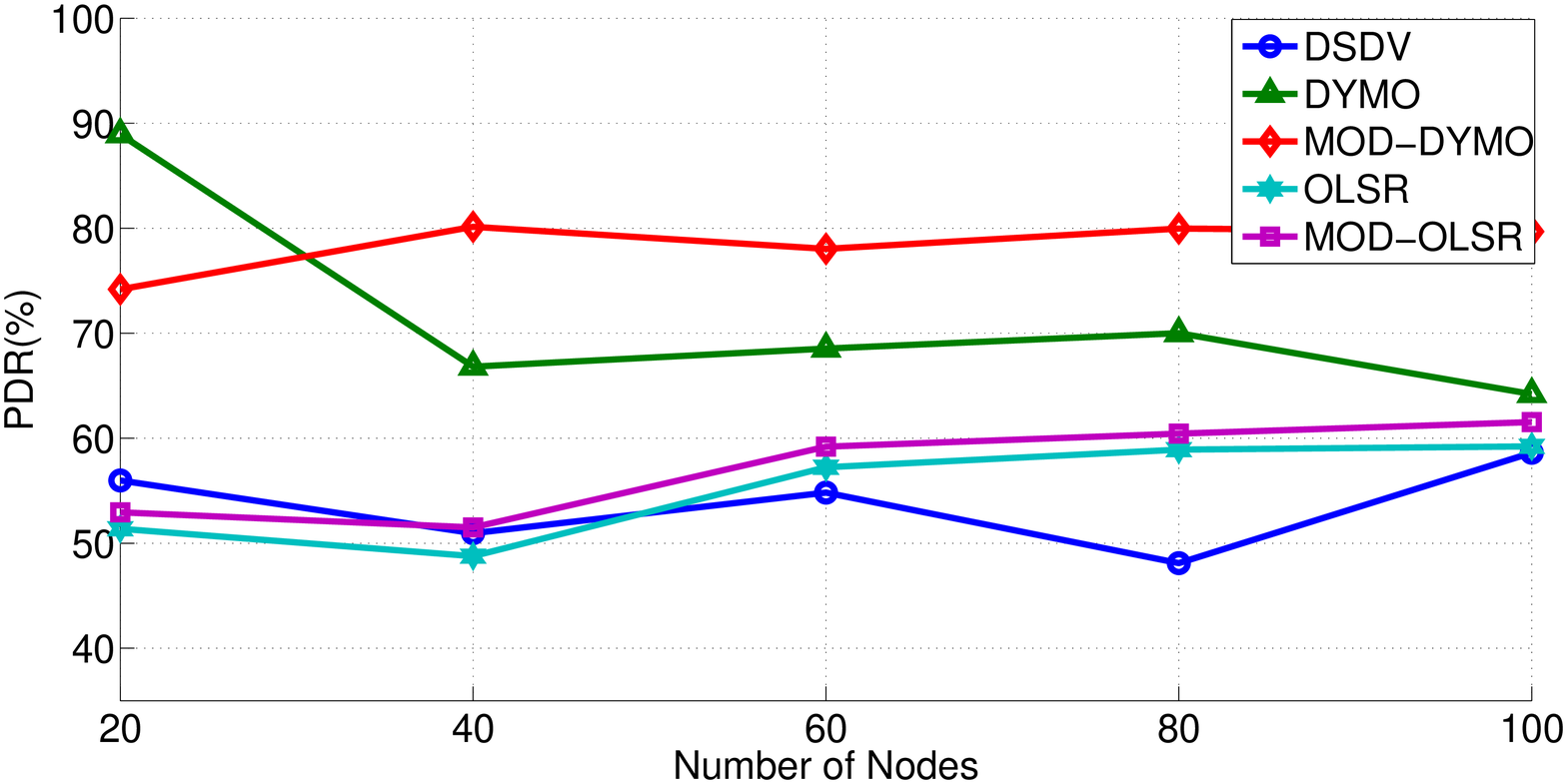}}
 \caption{PDR vs Scalability}
\end{figure}

In Fig. 4(b), we simulate PDR against node density. In Fig. 4(b), MOD-DYMO and DYMO sustain higher PDR than DSDV, MOD-OLSR and OLSR. The main reason of high PDR value is due to its reactive nature, because reactive protocols do not require more computation for route discovery. That is why MOD-DYMO performs well as compared to all other routing protocols due to reduced network diameter. Whereas, DSDV attains high value in low scalability and in medium scalability it comes down, however, as nodes density becomes high it shows high value due to more routing packets. MOD-OLSR and OLSR in low scalability show low data delivery value due to low optimization but as node density increases its PDR value also increases because high optimization of MPRs.
\begin{figure}[h]
  \centering
 \subfigure[Average PDR vs Number of connections]{\includegraphics[height=3 cm,width=4.3 cm]{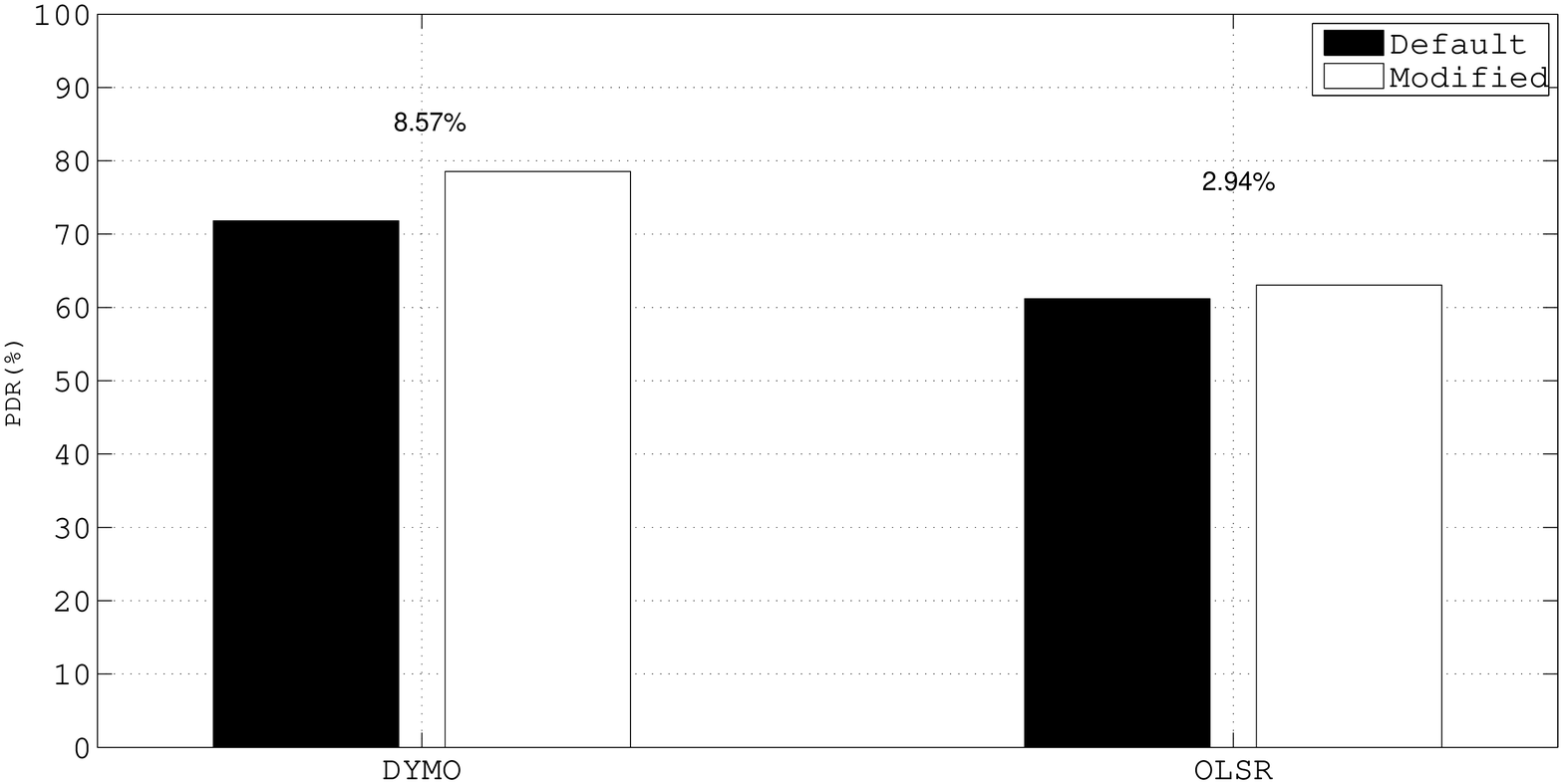}}
 \subfigure[Average PDR vs Node Density]{\includegraphics[height=3  cm,width=4.3 cm]{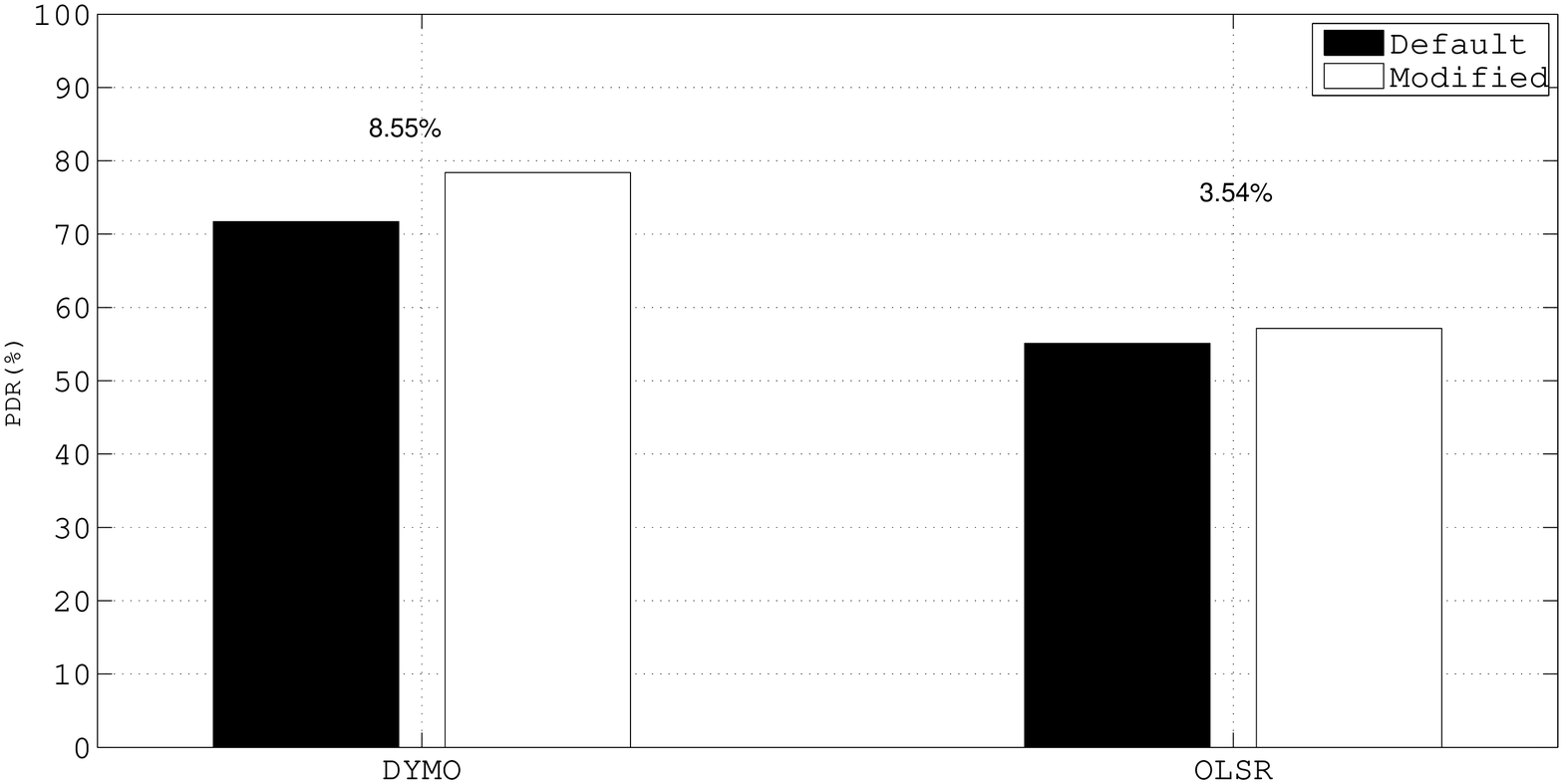}}
 \caption{Bar chart of PDR}
\end{figure}

In Fig. 5, we observe that enhanced versions of routing protocols perform better than default one. MOD-DYMO outperforms DYMO due to decrease in $Route~Request~wait~time$ from $1000$ to $600$ seconds and $network~diameter$ from $10$ to $30$ hops. Whereas, MOD-OLSR shows good results than OLSR due to decrement in intervals of updates; periodic and trigger.

Link duration and path stability in MOD-DYMO and DYMO is greater than DSDV, MOD-OLSR and OLSR because of high value of PDR. The main reason is less drop of packets causes more PDR and the link duration so as the path stability. DSDV, MOD-OLSR and OLSR also have good value of link duration and path stability but not as good as DYMO has.

\subsection{AE2ED}
Overall Delay of packet generation at the source and arrival at destination is known as AE2ED. Fig. 6(a) shows AE2ED against number of connections. OLSR and MOD-OLSR show highest value of AE2ED than other routing protocols; DSDV, DYMO and MOD-DYMO. Two main reasons, firstly proactive routing protocols have more AE2ED because before data transmission, they need to calculate routing tables. Secondly, generation of Hello and TC messages for checking the link and computing MPRs that causes more delay, therefore MOD-OLSR has less value than OLSR due to decrease in $Hello~and~TC~message~intervals$. DSDV attains high value than MOD-DYMO and DYMO, but less than MOD-OLSR and OLSR. DSDV has two main reasons for its high value, first proactive nature and second, the selection of best routes creates delay in advertising routes. DYMO has less AE2ED than DSDV, MOD-OLSR and OLSR, because it uses Expanding Ring Search (ERS) algorithm that reduces AE2ED. While MOD-DYMO performs better than DYMO due to decrease in request wait time.
\begin{figure}[h]
  \centering
 \subfigure[AE2ED vs Number of connections]{\includegraphics[height=3 cm,width=4.3 cm]{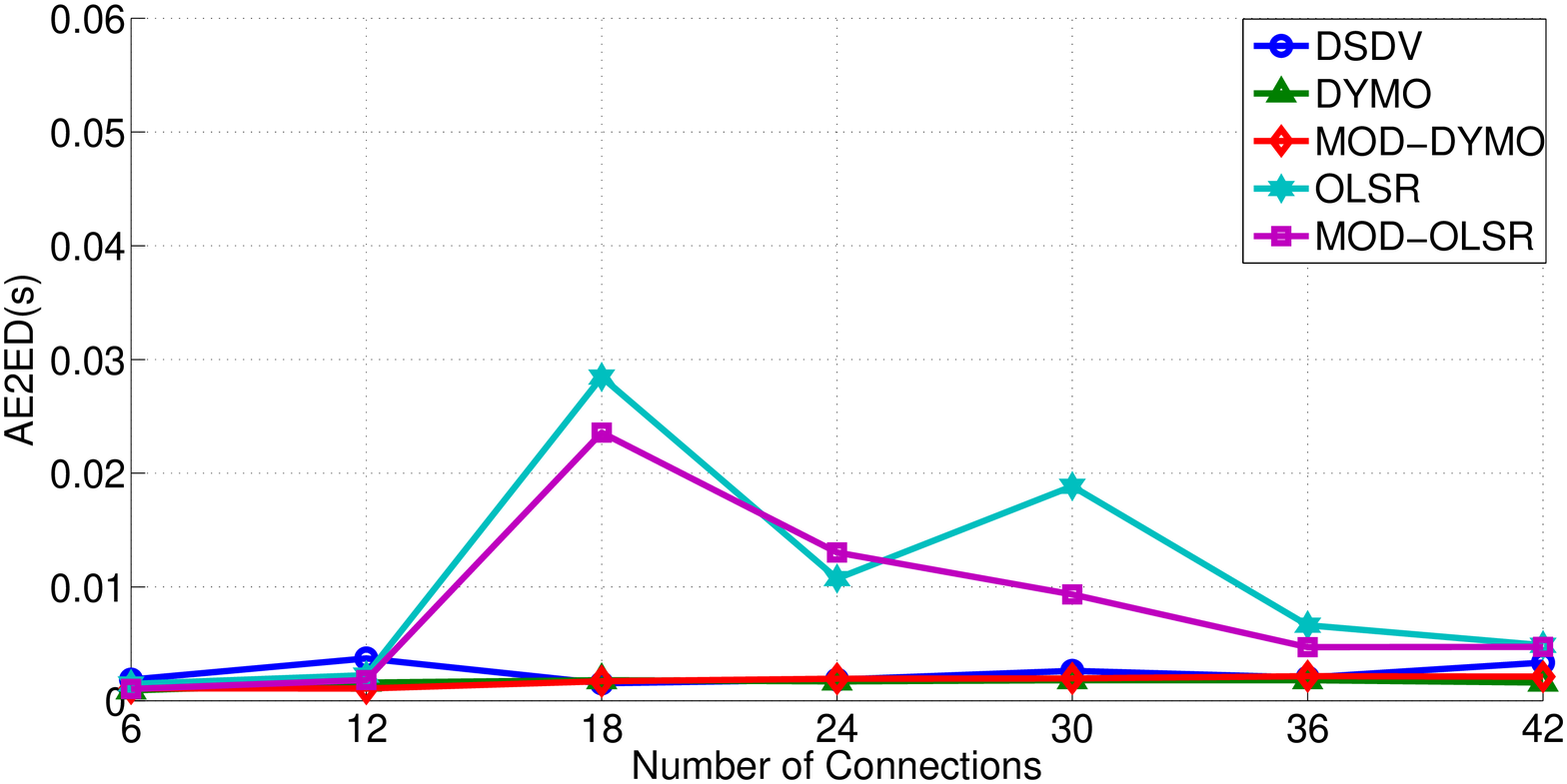}}
 \subfigure[AE2ED vs Node Density]{\includegraphics[height=3  cm,width=4.3 cm]{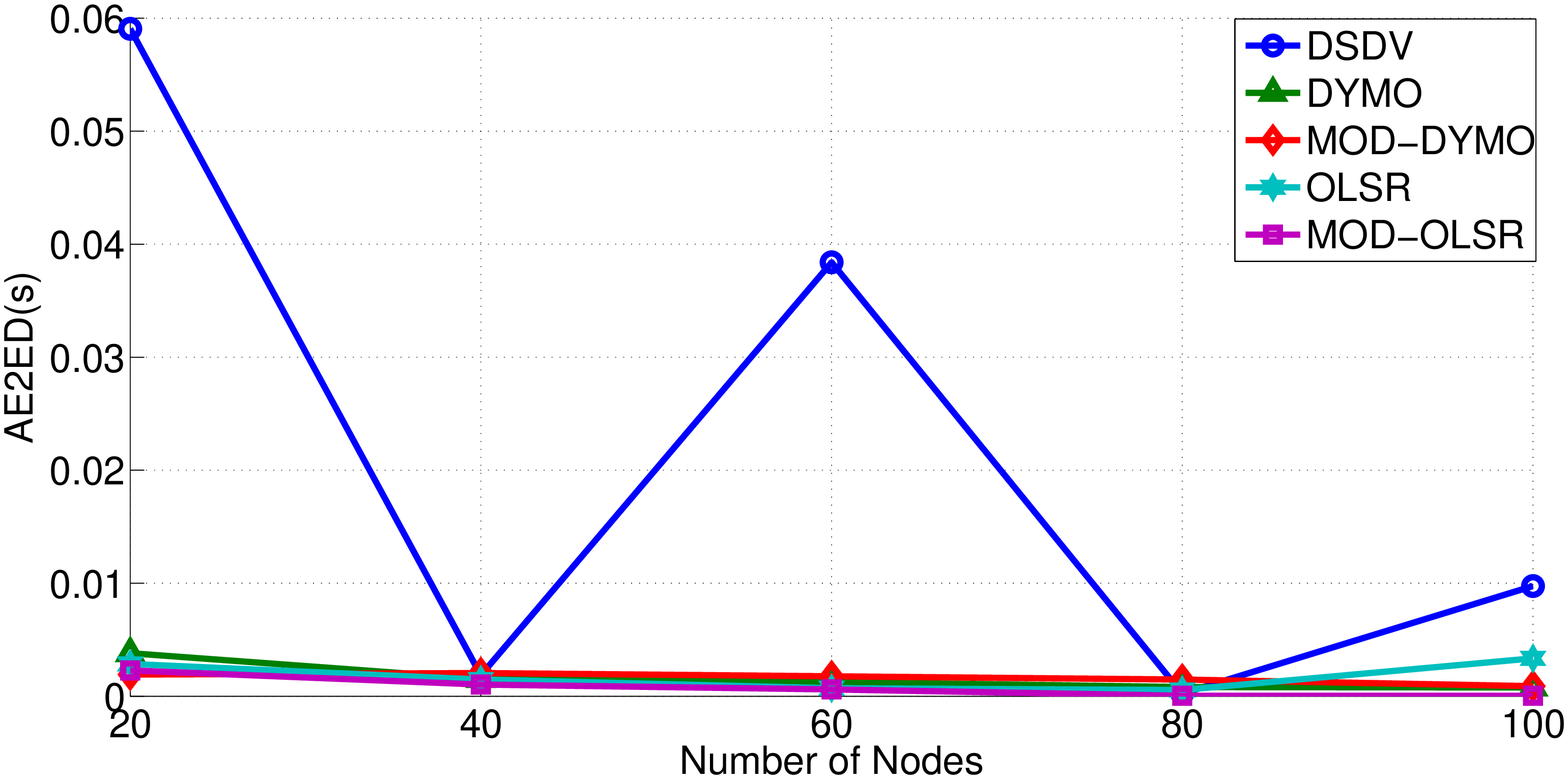}}
 \caption{AE2ED vs Scalability}
\end{figure}

In Fig. 6(b), we calculate AE2ED against node density. MOD-OLSR and OLSR has less delay because there is generation of Hello and TC messages for the link sensing and computing MPRs that causes reduction in delay. In low scalability, MOD-OLSR shows high value of AE2ED than OLSR because of decrement in $Hello~and~TC~message~intervals$. Overall, AE2ED of DSDV is very high, while in medium scalability, it shoots down with less AE2ED. MOD-DYMO and DYMO show less and almost decreased AE2ED. In medium scalability, DYMO and OLSR perform well by showing same delay.
\begin{figure}[h]
  \centering
 \subfigure[AE2ED vs Number of connections]{\includegraphics[height=3 cm,width=4.3 cm]{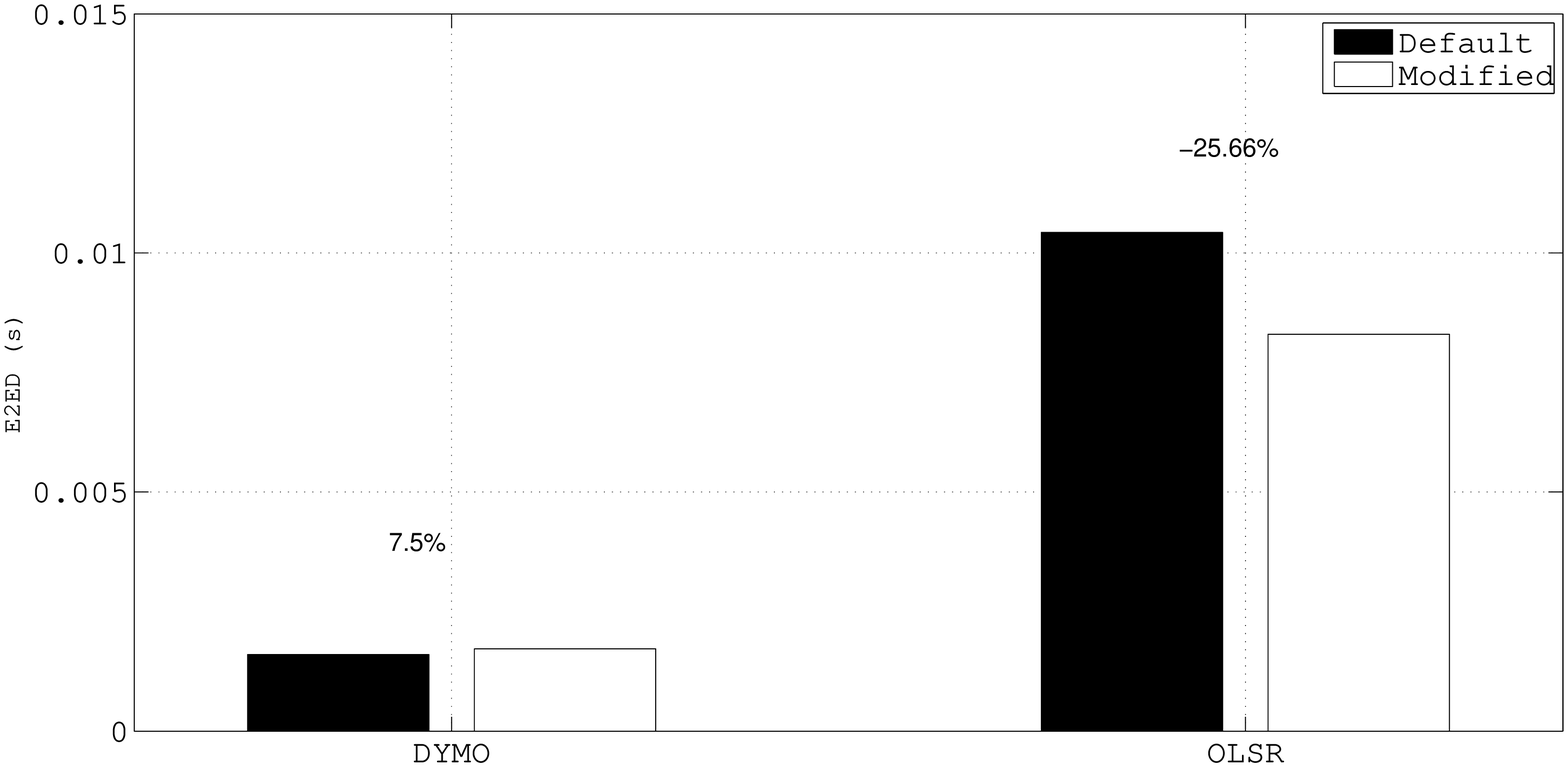}}
 \subfigure[AE2ED vs Node Density]{\includegraphics[height=3  cm,width=4.3 cm]{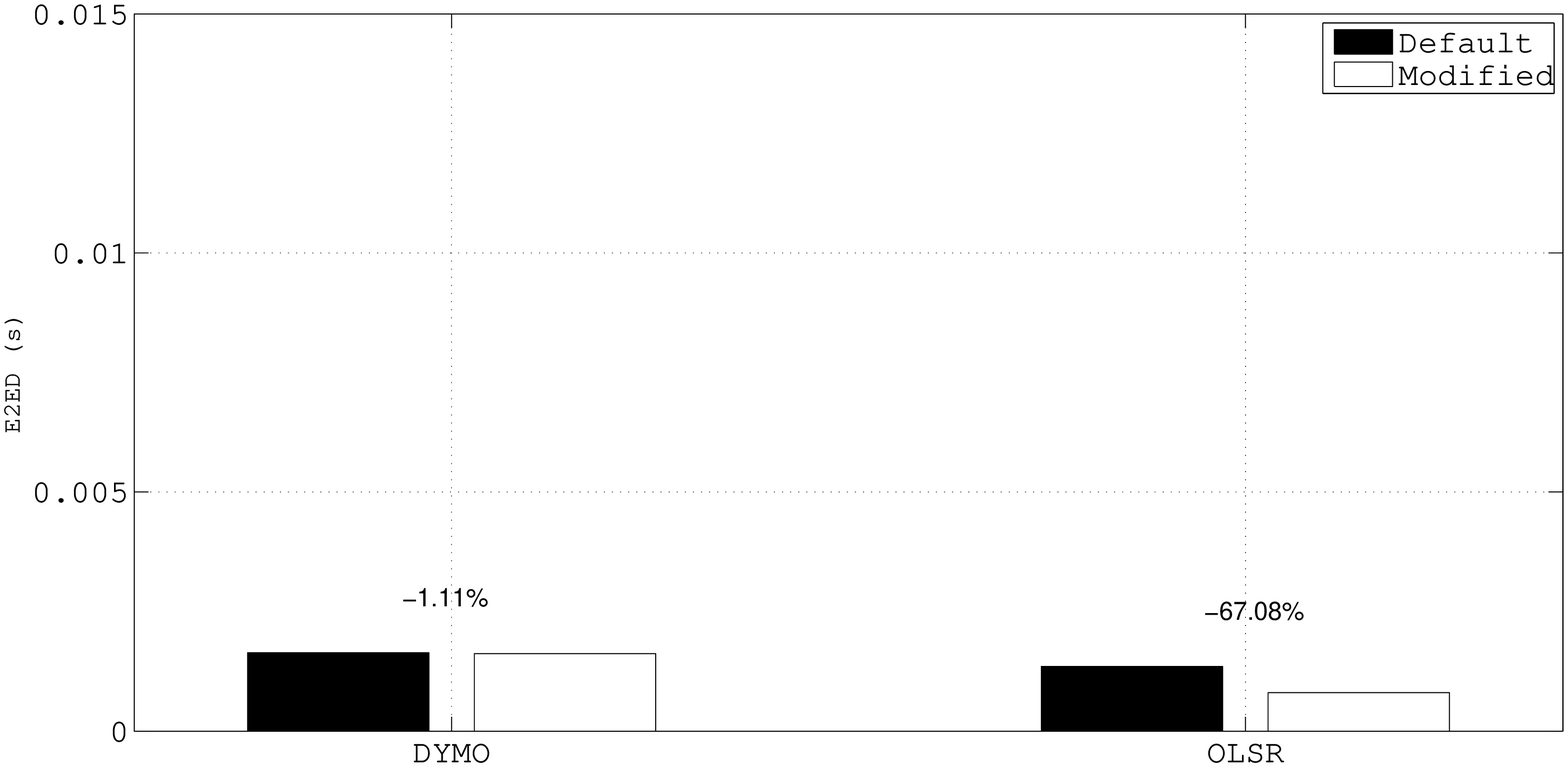}}
 \caption{Bar chart of AE2ED }
\end{figure}

In Fig. 7, DYMO outperforms MOD-DYMO due to decrease in $Route~Request~wait~time$ from $1000$ to $600$ seconds, that causes to decrease in delay. Whereas, MOD-OLSR shows less value of AE2ED than OLSR due to decrement in intervals of updates; TC and Hello Messages.

MOD-DYMO and DYMO sustain less value for AE2ED, due to showing best value of link duration and path stability. While, DSDV, MOD-OLSR and OLSR doe not have good value of link duration and path stability like DYMO, because proactive routing protocols have different link sensing updates.

\subsection{NRO}
The number of routing packets transmitted per data packet delivered to the destination is termed as NRO. From Fig. 8(a), it is observed that NRO of DYMO and MOD-DYMO is larger than both proactive routing protocols; DSDV, MOD-OLSR and OLSR. The reason is that DYMO is reactive in nature; ERS algorithm instead of LLR that is efficient for less delay and high NRO than rest of routing protocols. MOD-OLSR and OLSR has higher value of NRO than DSDV due to one reason that there is generation of Hello and TC messages for checking the link and computing MPRs that causes reduction in delay and increase in NRO. While OLSR has less value of NRO than MOD-OLSR due to decrease in $Hello~and~TC~message-intervals$. DSDV generates less NRO but in higher scalability DSDV sustains more NRO. The reason is that in high scalability, DSDV generates trigger updates and also periodic updates causes more NRO.

\begin{figure}[h]
  \centering
 \subfigure[NRO vs Number of connections]{\includegraphics[height=3  cm,width=4.3 cm]{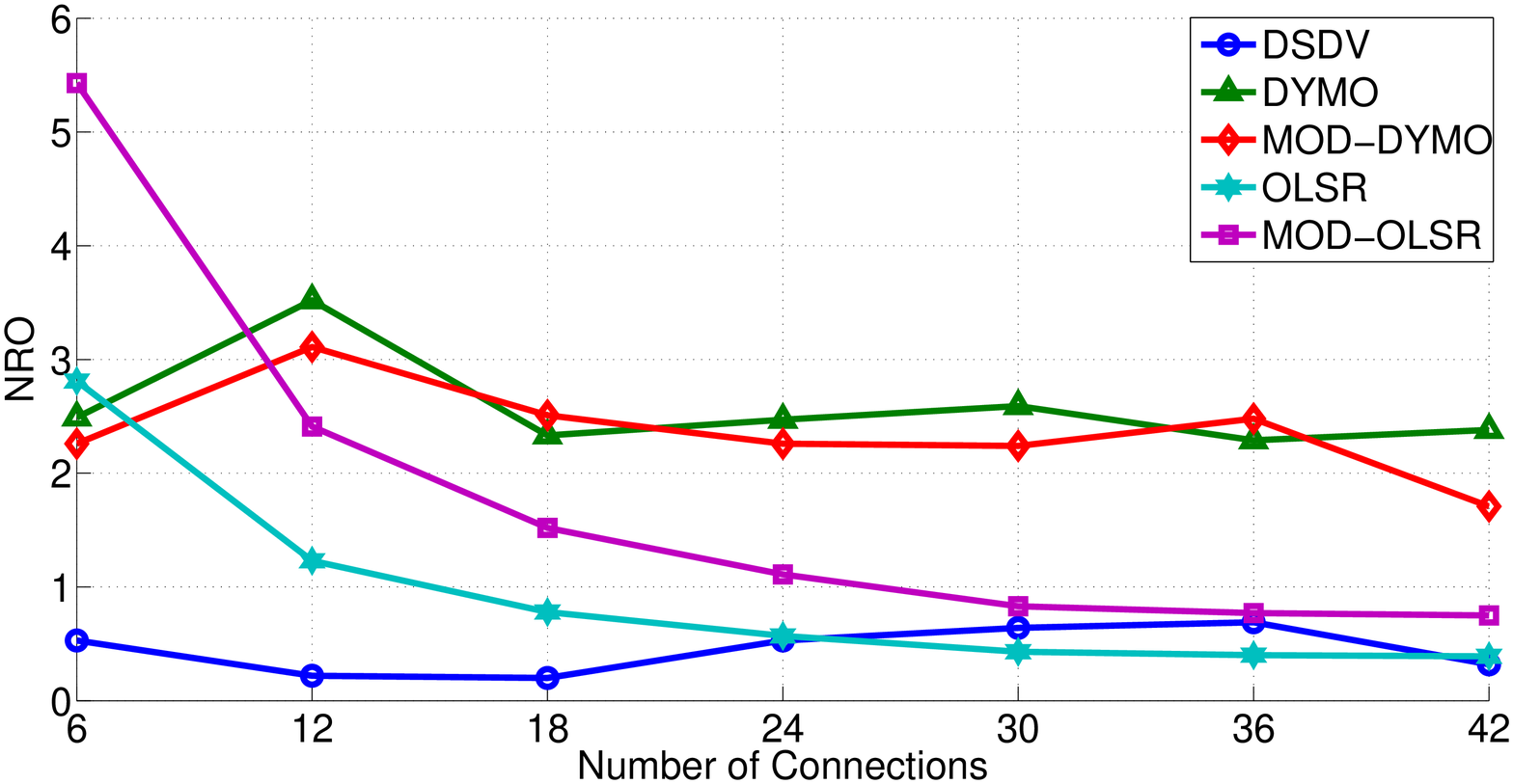}}
 \subfigure[NRO vs Node Density]{\includegraphics[height=3  cm,width=4.3 cm]{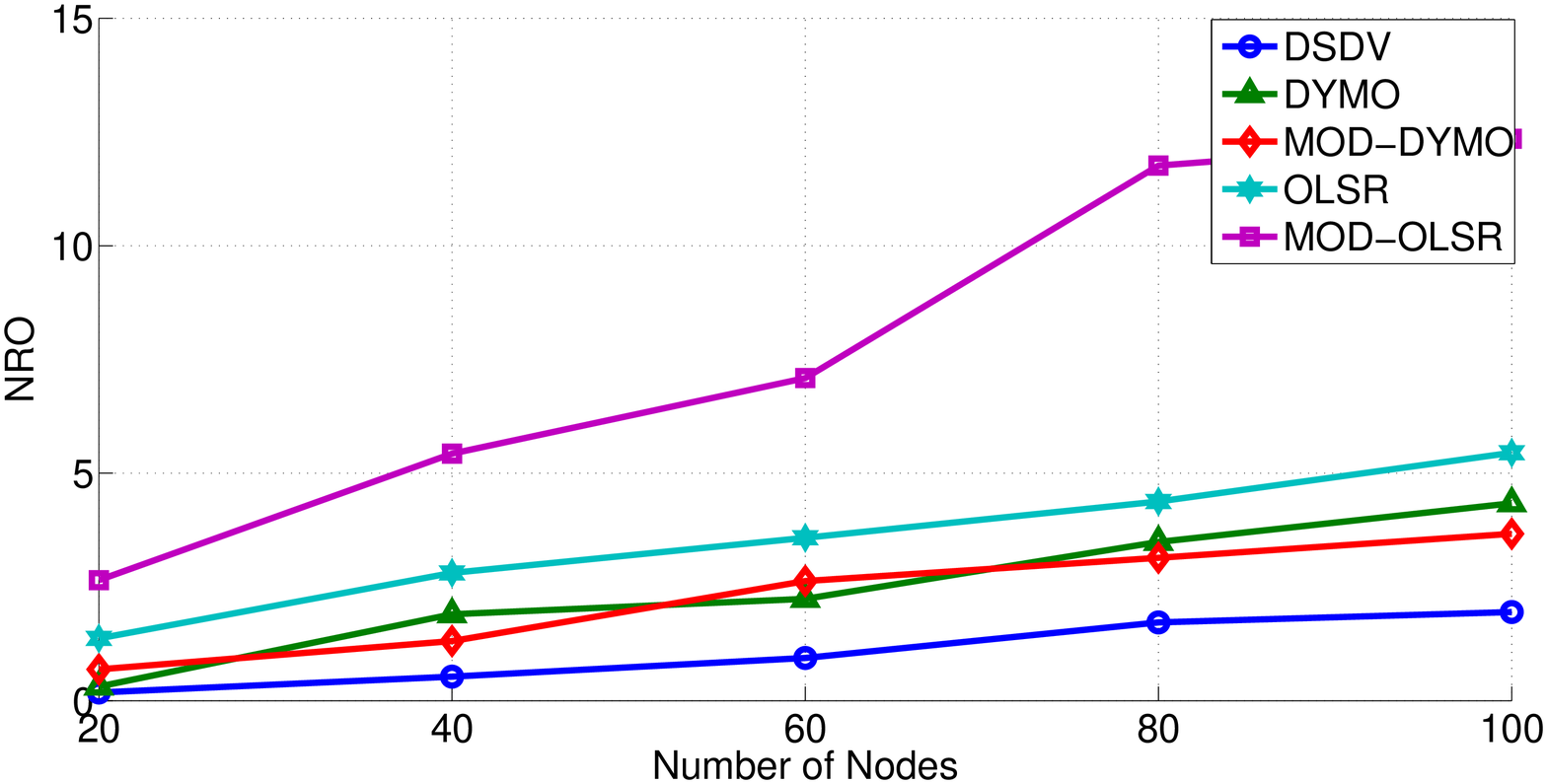}}
 \caption{NRO vs Scalability}
\end{figure}

In Fig. 8(b), MOD-OLSR has highest value of NRO but as node density increases its NRO increases with rising slope, due to more trigger updates. While OLSR sustains less value of NRO than MOD-OLSR and greater than DSDV and DYMO, due to short intervals of trigger updates; Hello and TC messages. DYMO attains high value of NRO but not greater than OLSR, due to route discovery on-demand. DSDV has less value of NRO but as node density increases, due to generation of more periodic and trigger updates.

\begin{figure}[h]
  \centering
 \subfigure[Average NRO vs Number of connections]{\includegraphics[height=3 cm,width=4.3 cm]{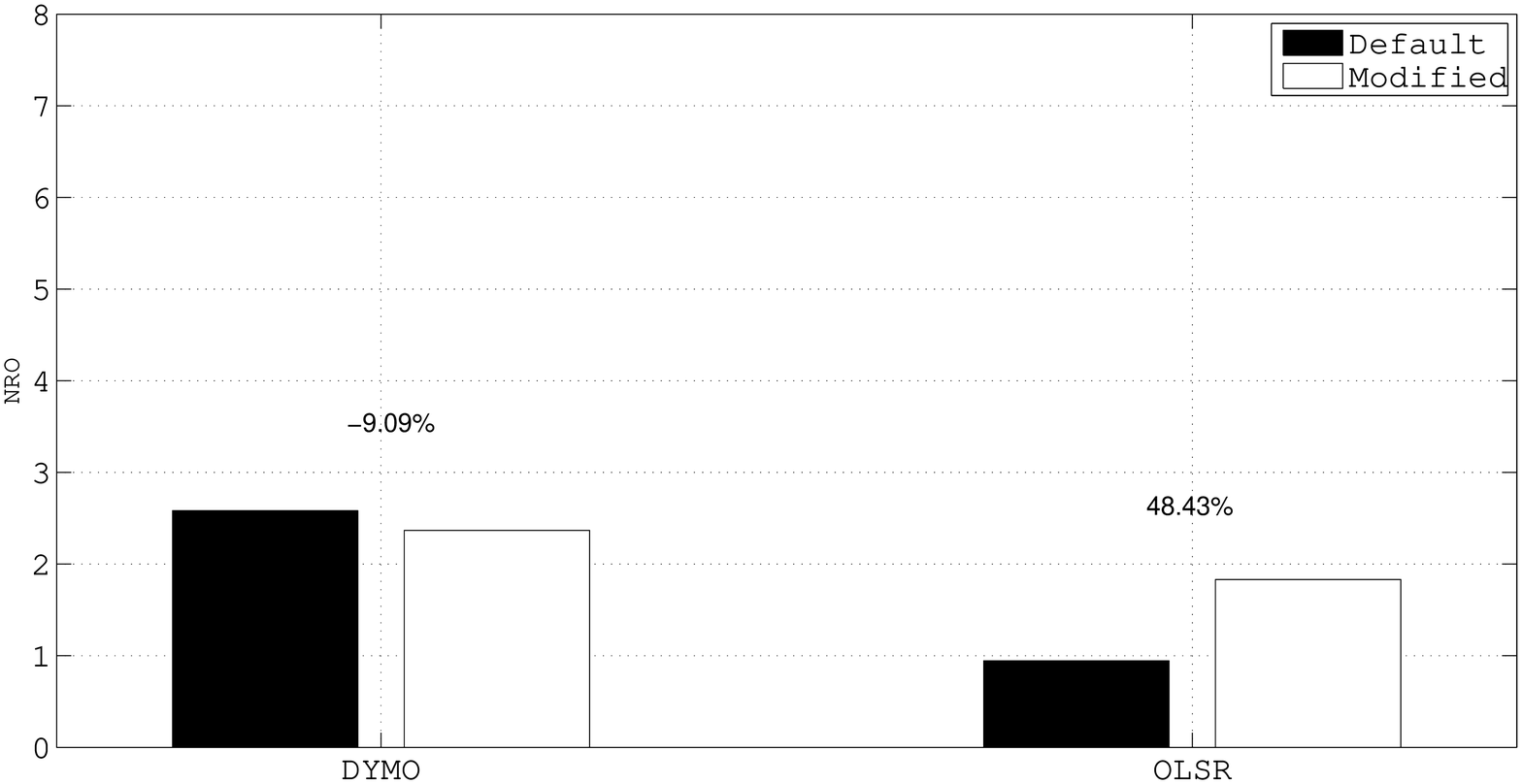}}
 \subfigure[Average NRO vs Node Density]{\includegraphics[height=3  cm,width=4.3 cm]{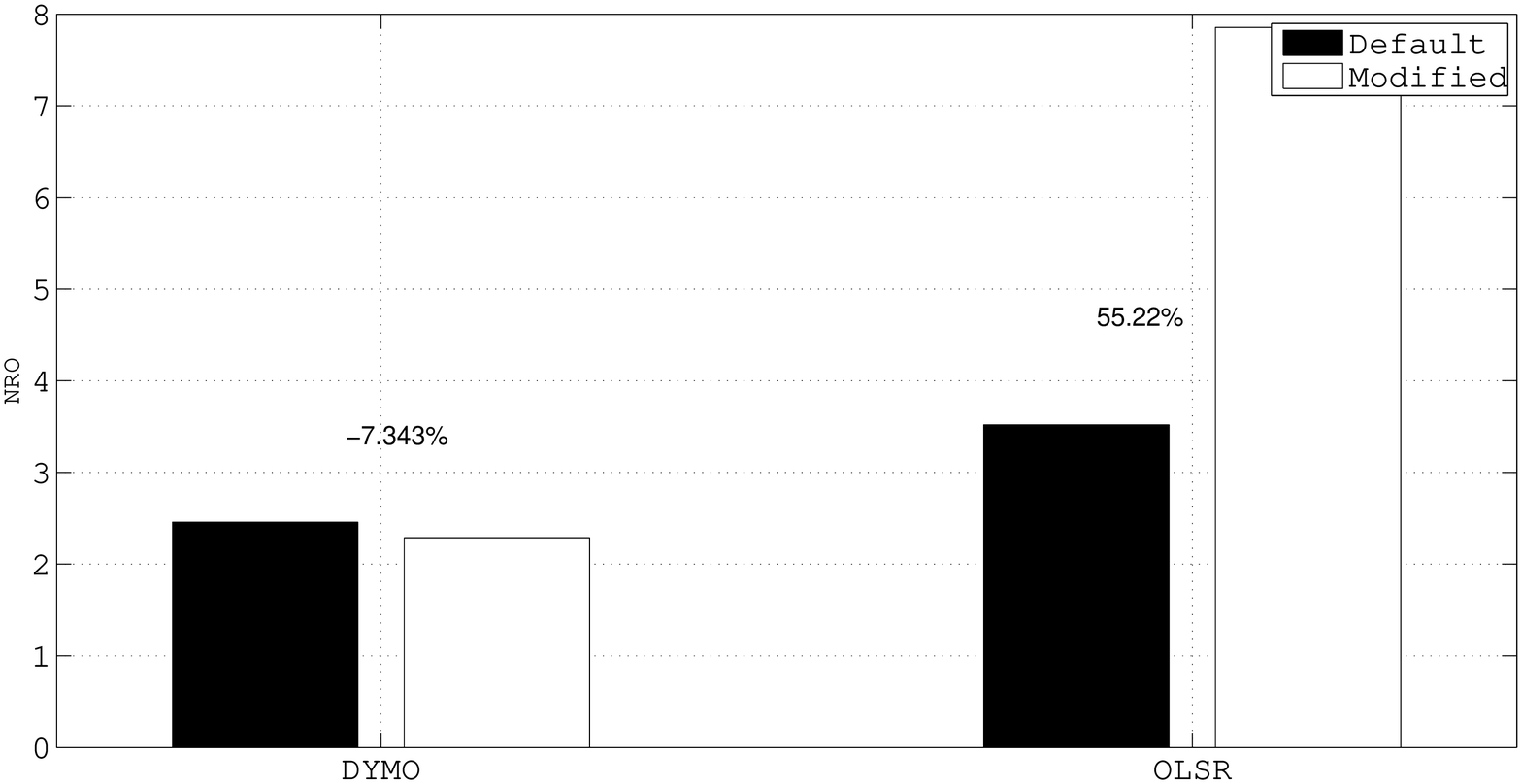}}
 \caption{Bar chart of NRO}
\end{figure}

In Fig. 9, we observe that default routing protocols sustain less value of NRO than modified. MOD-DYMO outperforms DYMO due to decrease in $Route~Request~wait~time$ from $1000$ to $600$ seconds and $network~diameter$ from $10$ to $30$ hops. Whereas, OLSR shows good results than MOD-OLSR due to decrement in intervals of TC and Hello Message.

DYMO and MOD-DYMO sustain link duration for longer time and good value of path stability, that will cause high value of NRO. OLSR and MOD-OLSR have better link duration than DSDV but not than DYMO and MOD-DYMO due to its MPRs mechanism.

\section{Conclusion}
In this paper, we simulate and compare routing protocols, both default and enhanced versions under the performance parameters; PDR, AE2ED and NRO. From the results, it is concluded that DYMO performs better than all routing protocols but not better than MOD-DYMO, in terms of PDR and AE2ED at the cost of high value of NRO. Whereas, DSDV performs better for NRO and less value of AE2ED in terms of number of connections and node density. MOD-OLSR and OLSR sustain average value for PDR and less AE2ED at the cost of very high value of NRO. MOD-DYMO and DYMO outperform DSDV, MOD-OLSR and OLSR in terms of link duration and path stability at the cost of high value of NRO.

\end{document}